\documentstyle[pre,aps]{revtex}

\begin{document}

\draft

\title{Anomalous Transient Current in Nonuniform Semiconductors} 

\author{V.I. Yukalov$^{1,2}$, E.P. Yukalova$^{3,4}$, and M.R. Singh$^3$} 

\address{\it $^1$ Centre for Interdisciplinary Studies in Chemical Physics \\
University of Western Ontario, London, Ontario N6A 3K7, Canada \\
$^2$Bogolubov Laboratory of Theoretical Physics\\
Joint Institute for Nuclear Research, Dubna 141980, Russia \\
$^3$Department of Physics and Astronomy\\
University of Western Ontario, London, Ontario N6A 3K7, Canada \\
$^4$ Laboratory of Computing Techniques and Automation \\
Joint Institute for Nuclear Research, Dubna 141980, Russia}

\maketitle

\begin{abstract}

Nonequilibrium processes in semiconductors are considered with highly 
nonuniform initial densities of charge carriers. It is shown that there 
exist such distributions of charge densities under which the electric 
current through a sample displays quite abnormal behaviour flowing
against the applied voltage. The appearance of this negative 
electric current is a transient phenomenon occurring at the initial stage 
of the process. After this anomalous negative fluctuation, the electric 
current becomes normal, i.e. positive as soon as the charge density 
becomes more uniform. Several possibilities for the practical usage of 
this effect  are suggested.
\end{abstract}

\vspace{1cm}


\section{Introduction}

The study of electric processes in semiconductor materials plays an 
important role in understanding the physics of semiconductor devices as 
well as in their design and development [1,2]. One of the most difficult 
problems is the consideration of strongly nonequilibrium effects in
essentially nonuniform semiconductors. At the same time the latter can
display quite interesting specific features caused by nonuniform
distributions of charge carriers [3--6].

For instance, electric current through a semiconductor device can display
rather abnormal behaviour, with transient fluctuations corresponding to
the flow of the current against the applied voltage [5,6]. In the short
communications [5,6] a model case of a unipolar semiconductor was considered, 
with numerical analysis not including the relaxation parameters and diffusion 
coefficients.

The aim of the present paper is to study nonequilibrium processes, with a
strongly nonuniform initial distribution of charge carriers, for realistic
semiconductor materials. We consider the general case of a semiconductor
with two kinds of charge carriers, positive and negative. Numerical
analysis takes account of relaxation and diffusion effects. The total
current through a semiconductor device is considered, together with the
currents across the left and right surfaces of this device. And also the
role of the generation--recombination noise is analysed.

In Sec.2 we collect the main equations defining the problem and needed for 
the following analysis. An approximate analytical solution of these equations 
is presented in Sec.3, which permits to show explicitly the motion of charge 
carriers under an applied voltage. The conditions for the occurrence of the 
negative electric current, directed against this applied voltage, are derived 
in Sec.4. This anomalous transient effect is illustrated by numerical 
solutions. Sec.5 contains conclusion and discussion with several 
suggestions for the possible practical usage of the considered effect. 
Appendix contains the proof that the approximate regular solution 
asymptotically coincides with the regular stable solution of the exact 
equations.

\section{Basic Equations}

The charge carriers are characterized by the densities $\;\rho_1>0\;$ and 
$\;\rho_2<0\;$ that are functions of the space position 
$\;\stackrel{\rightarrow}{r}\;$ and time $\;t\;$, i.e. 
$\;\rho_i=\rho_i(\stackrel{\rightarrow}{r},t)\;$ with $\;i=1,2\;$. Carrier 
transport can be described in terms of a semiclassical approach called the 
drift--diffusion approximation which is the basis of the majority of 
semiconductor device models [1,2]. This approach is, of course, 
phenomenological since it does not concern the microscopic derivation of 
the used parameters, such as mobilities or relaxation widths. The values 
of these parameters depend on a number of different underlying causes. 
For instance, the lattice structure of the considered semiconductor is 
important for defining the values of these parameters. Thus, the 
scattering of carriers on phonons influences both mobilities as well as 
damping parameters. However, the calculation of such parameters is a 
separate problem that is not the aim of the present paper. In the 
semiclassical approach all parameters are assumed to be given a priori. 
With the given phenomenological parameters, the drift--diffusion 
approximation is known to give a very good description of realistic 
semiconductor devices, which explains why this approximation is so widely 
used [1,2]. The first set of equations in this approach consists of the set 
of continuity equations
\begin{equation}
\frac{\partial\rho_i}{\partial t} +\stackrel{\rightarrow}{\nabla}
\cdot\stackrel{\rightarrow}{j}_i + \frac{\rho_i}{\tau_i} =\xi_i
\end{equation}
for each kind of carriers. The relaxation term, with the relaxation time 
$\;\tau_i\;$, is taken in the simplest form since, as will be clear from 
what follows, it does not play an essential role in transient processes 
occurring at times $\;t\ll\tau_i\;$. The right--hand side of (1) is the
generation--recombination noise [7] which is always present in semiconductor 
devices. Not to overcomplicate the problem we do not include other types
of noise [7--9] assuming that they are of less importance.

Another equation is a Maxwell equation 
\begin{equation}
\varepsilon\stackrel{\rightarrow}{\nabla} \cdot \stackrel{\rightarrow}{E}
 = 4\pi(\rho_1 +\rho_2) \; ,
\end{equation}
with the dielectric permittivity $\;\varepsilon\;$.

The electric--current density in (1) is
\begin{equation}
\stackrel{\rightarrow}{j}_i =
\mu_i\rho_i\stackrel{\rightarrow}{E} -
D_i\stackrel{\rightarrow}{\nabla}\rho_i ,
\end{equation}
where $\;\mu_1>0\;$ and $\;\mu_2<0\;$ are the carrier mobilities and 
$\;D_i\equiv(\mu_i/e_i)k_BT\;$ are the diffusion coefficients in which 
$\;e_1>0\;$ and $\;e_2<0\;$ are the carrier charges, $\;k_B\;$ is the 
Boltzmann constant, and $\;T\;$ is temperature. The first and second 
terms in (3) are the drift and diffusion current densities, respectively. 
Adding here the displacement current, one has the total current density
\begin{equation}
\stackrel{\rightarrow}{j}_{tot} =
\stackrel{\rightarrow}{j}_1 + \stackrel{\rightarrow}{j}_2 + 
\frac{\varepsilon}{4\pi}
\frac{\partial\stackrel{\rightarrow}{E}}{\partial t}\; .
\end{equation}

If the device is biased with an externally applied voltage $\;V_0\;$ 
along a path $\;\stackrel{\rightarrow}{l}\;$, then
\begin{equation}
\int_{\stackrel{\rightarrow}{l}}\stackrel{\rightarrow}{E}
(\stackrel{\rightarrow}{r},t)d \stackrel{\rightarrow}{l} = V_0 .
\end{equation}
For concreteness, we consider a positive bias, that is, $\;V_0>0\;$. It is 
assumed that metal contacts supplying the external voltage are nondamaging, 
that is, do not induce in their vicinity incubation effects.

Consider a plane device of the width $\;L\;$ and area $\;A\;$. Then instead of
$\;\stackrel{\rightarrow}{r}\;$ we are to deal with one space variable 
$\;x\in[0,L]\;$.

An important characteristic is the transit time
\begin{equation}
\tau_0\equiv\frac{L^2}{\mu V_0} ,
\end{equation}
where $\;\mu=\min\{\mu_1,|\mu_2|\}\;$. Usually, $\;\mu_1<|\mu_2|\;$.
This time is to be compared with the relaxation times $\;\tau_i\;$. It 
is often more convenient to deal with inverse times called widths. For 
instance,
\begin{equation}
\gamma_i\equiv\frac{1}{\tau_i} ,
\end{equation}
with $\;i=1,2\;$, define the relaxation widths.

Another convenience is to deal with dimensionless quantities, which will 
be done in what follows. To return to dimensional quantities, we shall 
imply that $\;x\;$ is measured in units of $\;L\;$; $\;t\;$ and $\;\tau_i\;$,
in units of $\;\tau_0\;$; and other physical quantities, in the 
corresponding units listed below:
$$ D_0\equiv \mu V_0 , \qquad E_0\equiv\frac{V_0}{L} , \qquad Q_0\equiv
\varepsilon AE_0 , $$
\begin{equation}
\rho_0\equiv\frac{Q_0}{AL} , \qquad \xi_0\equiv\frac{\rho_0}{\tau_0},
\qquad J_0\equiv\frac{Q_0L}{\tau_0} .
\end{equation}

For the case considered, Eq.(1) reduces to
\begin{equation}
\frac{\partial\rho_i}{\partial t} +\mu_i\frac{\partial}{\partial x}
\left (\rho_iE\right ) - D_i\frac{\partial^2\rho_i}{\partial x^2} +
\gamma_i\rho_i =\xi_i ,
\end{equation}
and Eq.(2) becomes
\begin{equation}
\frac{\partial E}{\partial x} =4\pi(\rho_1 +\rho_2) ,
\end{equation}
where
\begin{equation}
0 < x < 1 , \qquad t > 0 .
\end{equation}
Condition (5) for an applied voltage reads
\begin{equation}
\int_0^1E(x,t)dx = 1 .
\end{equation}
These equations are to be supplemented by initial conditions
\begin{equation}
\rho_i(x,0) =f_i(x)  \qquad (i=1,2) .
\end{equation}

Electric field can be expressed, from Eq.(10), as a functional
\begin{equation}
E(x,t) = 1 +4\pi\left [ Q(x,t) -\int_0^1Q(x,t)dx\right ]
\end{equation}
of the charge densities, so that
\begin{equation}
Q(x,t) =\int_0^x\left [ \rho_1(x',t) +\rho_2(x',t)\right ]dx' .
\end{equation}
And the total density of current (4) can be written as
\begin{equation}
j_{tot} =\mu_1\rho_1 E - D_1\frac{\partial\rho_1}{\partial x} +
\mu_2\rho_2 E - D_2\frac{\partial\rho_2}{\partial x} + 
\frac{1}{4\pi}\frac{\partial E}{\partial t} \; ,
\end{equation}
where $\;j_{tot}=j_{tot}(x,t)\;$.

The quantities that can be measured and that we are going to study in what
follows are the current across the left surface
\begin{equation}
J(0,t)\equiv j_{tot}(0,t) \; ,
\end{equation}
the current across the right surface
\begin{equation}
J(1,t) \equiv j_{tot}(1,t) \; ,
\end{equation}
and the total current through the device
\begin{equation}
J(t) = \int_0^1 j_{tot}(x,t) dx \; .
\end{equation}
The latter, using Eqs. (16) and (12), can be presented as
$$
J(t) = \int_0^1
\left [ \mu_1\rho_1(x,t) + \mu_2\rho_2(x,t)\right ] E(x,t)dx +
$$
\begin{equation}
+ D_1 \left [ \rho_1(0,t) - \rho_1(1,t)\right ] +
D_2 \left [ \rho_2(0,t) - \rho_2(1,t)\right ] \; .
\end{equation}

Our aim is to study the peculiarities in the time dependence of the
electric current when the initial conditions (13) correspond to a strongly
nonuniform charge distribution. Such nonuniform distributions can be
prepared in different ways. For example, one can organize a nonuniform
distribution in the process of growing of a semiconductor sample. Another
way is to irradiate semiconductor by narrow laser beams [10]. One more
possibility is by forming heavily doped layers by ion irradiation [11].
This method makes it possible to form narrow layers of positive carriers
with a density of $\;10^{20}cm^{-3}\;$.

\section{Carrier Densities}

To understand better the physics of processes resulting from a nonuniform 
initial distribution of charge carriers, it would be useful to find 
an analytical, though approximate, solution to the system of equations 
(9) and (10). This can be done by means of the method of scale 
separation [12,13], whose mathematical foundation is based on the
Krylov--Bogolubov averaging method [14].

The first step in the method of scale separation [12,13] is to classify 
the solutions onto fast and slow. In our case this can be done as follows. 
The electric field, as is seen from Eq. (14), is the functional of 
the charge densities, averaging the latter over the space variable $\;x\;$.
This results in that $\;E\;$ varies in space slower than $\;\rho_i\;$. On 
the other hand, the voltage integral (12) shows that the electric field, 
being averaged over space, does not depend on time. This means that 
$\;E\;$ can be treated as a slow function in time. Therefore, the electric 
field $\;E\;$ can be regarded as a slow solution, as compared to the 
charge densities $\;\rho_i\;$, with respect to both space and time. This 
permits us to consider the equation (9) for a fast solution $\;\rho_i\;$ 
keeping there $\;E\;$ as a space--time quasi--integral. After solving 
Eq. (9), the found $\;\rho_i\;$ is to be substituted into Eq. (14) giving 
an equation for $\;E\;$ which can be solved iteratively.

In solving Eq. (9), it is convenient to continue $\;\rho_i\;$ outside the 
region of $\;x\in(0,1)\;$ by defining $\;\rho_i\;$ as zero for $\;x < 
0\;$ and $\;x > 1\;$. Then we may invoke the Fourier transforms with 
respect to $\;x\;$. Finally, we obtain
\begin{equation}
\rho_i(x,t) =\rho_i^{reg}(x,t) +\rho_i^{ran}(x,t) ;
\end{equation}
the first term being the regular solution
\begin{equation}
\rho_i^{reg}(x,t) =\int_{-\infty}^{+\infty}G_i(x-x',t)f_i(x')dx'
\end{equation}
induced by the initial condition (13), while the second term being the 
random solution
\begin{equation}
\rho_i^{ran}(x,t) =\int_0^t\int_{-\infty}^{+\infty} 
G_i(x-x',t-t')\xi_i(x',t')dx'dt'
\end{equation}
generated by the noise. The Green function in Eqs. (22) and (23) is
\begin{equation}
G_i(x,t) =\frac{1}{2\pi}\int_{-\infty}^{+\infty}\exp\{ ikx 
-i\omega_i(k)t\} dk
\end{equation}
with the spectrum
\begin{equation}
\omega_i(k) =\mu_iEk- iD_ik^2 -i\gamma_i .
\end{equation}
Function (24) has the properties
$$ G_i(x,0) =\delta(x) , \qquad \int_{-\infty}^{+\infty}G_i(x,t)dx = 
e^{-\gamma_it}  . $$
In the case considered, the integration in (24) can be realized 
explicitly resulting in
\begin{equation}
G_i(x,t) =\frac{1}{2\sqrt{\pi D_it}}\exp\left\{ 
-\frac{(x-\mu_iEt)^2}{4D_it} -\gamma_it\right\} .
\end{equation}

As the initial condition in Eq. (13) it is reasonable to accept the 
physically realistic case of the Gaussian distribution
\begin{equation}
f_i(x) =\frac{Q_i}{Z_i}\exp\left\{-\frac{(x-a_i)^2}{2b_i}\right\} ,
\end{equation}
in which $\;0<a_i<1\;$ and
$$ Q_i=\int_0^1f_i(x)dx , \qquad 
Z_i=\int_0^1\exp\left\{-\frac{(x-a_i)^2}{2b_i}\right\} dx . $$
With the initial condition (27), the regular solution (22) becomes
\begin{equation}
\rho^{reg}_i(x,t)=\frac{Q_ib_i}{Z_i\sqrt{b_i^2+2D_it}}\exp\left\{
-\frac{(x-\mu_iEt-a_i)^2}{2b_i^2+4D_it}-\gamma_it\right\} .
\end{equation}

The regular and random solutions satisfy the initial conditions
\begin{equation}
\rho_i^{reg}(x,0) = f_i(x), \qquad \rho_i^{ran}(x,0) = 0 .
\end{equation}
The regular solution, as time increases, moves with the velocity 
$\;\mu_iE\;$, becomes wider and smaller, so that
\begin{equation}
\lim_{t\rightarrow\infty}\rho_i^{reg}(x,t)=0 .
\end{equation}
The behaviour of the random solution depends on that of the noise. It is 
customary to treat the latter as the white noise with the averaging 
properties
\begin{equation}
\langle\xi_i(x,t)\rangle = 0 , \qquad 
\langle\xi_i(x,t)\xi_j(x',t')\rangle =\gamma_{ij}\delta(x-x')\delta(t-t') .
\end{equation}
Accepting (31), one has
\begin{equation}
\langle\rho_i^{ran}(x,t)\rangle = 0 ,
\end{equation}
and, consequently,
\begin{equation}
\lim_{t\rightarrow\infty}\langle\rho_i(x,t)\rangle = 0 .
\end{equation}
Since the electric field, according to (14), is a linear functional in 
$\;\rho_i\;$, we find
\begin{equation}
\lim_{t\rightarrow\infty}\langle E(x,t)\rangle = 1 .
\end{equation}
Although the limiting values (33) and (34) have been obtained by analysing 
the approximate solutions, it is possible to show (see Appendix) that the 
limits (33) and (34) are stable stationary solutions of the exact equations.

The random solution (23) influences the electric current (2) through the 
correlator
\begin{equation}
\langle\rho_i^{ran}(x,t)\rho_j^{ran}(x',t)\rangle = 
\gamma_{ij}\int_0^tG_{ij}(x-x',t')dt' ,
\end{equation}
in which
\begin{equation}
G_{ij}(x-x',t)\equiv\int_{-\infty}^{+\infty}G_i(x-x'',t)G_j(x'-x'',t)dx'' .
\end{equation}
With the Green function (26), this gives 
\begin{equation}
G_{ij}(x,t) =\frac{1}{2\sqrt{\pi(D_i+D_j)t}}\exp\left\{ 
-\frac{(x-\mu_iEt+\mu_jEt)^2}{4(D_i+D_j)t} -(\gamma_i+\gamma_j)t\right\} .
\end{equation}

At large time, Eq.(37) decays by the law
\begin{equation}
G_{ij}(x,t)\simeq\frac{1}{2\sqrt{\pi(D_i+D_j)t}}\exp\left 
(-\gamma_{eff}t\right ) ,
\end{equation}
as $\;t\rightarrow\infty\;$, with the effective attenuation
$$ \gamma_{eff}\equiv\frac{(\mu_i-\mu_j)^2E^2}{4(D_i+D_j)} +\gamma_i 
+\gamma_j . $$
Consequently, the correlator (35) tends to a time constant.

At small time, one has
\begin{equation}
G_{ij}(x,t)\simeq\frac{1}{2\sqrt{\pi(D_i+D_j)t}}\exp\left\{ 
-\frac{x^2}{4(D_i+D_j)t}\right\} ,
\end{equation}
as $\;t\rightarrow  0\;$. Therefore, the correlator (35) behaves as
\begin{equation}
\langle\rho_i^{ran}(x,t)\rho_j^{ran}(x',t)\rangle\simeq
\frac{\gamma_{ij}\sqrt{t}}{2\sqrt{\pi(D_i+D_j)}}\exp\left\{ 
-\frac{(x-x')^2}{4(D_i+D_j)t}\right\} ,
\end{equation}
when $\;t\rightarrow 0\;$. Equation (40) shows that the influence of 
noise at small times is exponentially suppressed. This conclusion is of 
high importance for the following analysis.

\section{Electric Current}

We have now enough information about the behaviour of the system in order 
to answer the question: Is it possible that a negative electric current 
could appear, directed against the applied voltage?

The first evident necessary condition for such a possibility is the space 
nonuniformity of the charge densities. Really, if $\;\rho_i(x,t)\;$ is 
uniform in $\;x\;$, then from Eq. (20) it follows immediately that the 
electric current is the positively defined quantity 
$\;\mu_1\rho_1+\mu_2\rho_2>0\;$.

The properties of the carrier densities, studied in the previous section, 
are such that, even, if at the initial time $\;\rho_i(x,t)\;$ is 
nonuniform in space, it tends to become uniform with time. Consequently, 
if a negative electric current would appear, this could happen only at 
the initial stage of the process, when $\;t\ll 1\;$.

In this way, the occurrence of negative electric current, if any, can 
arise only as a principally transient effect, when the charge densities 
are yet nonuniform. The processes of diffusion and relaxation need some
time to make $\;\rho_i\;$ uniform. Therefore, there always can be found 
such a time $\;t\ll 1\;$ when diffusion and relaxation are yet not important. 
But these processes shorten the time of a negative--current fluctuation, 
if it appears. Thus, the conditions favoring the longer lifetime of such 
a fluctuation would be $\;D_i\ll 1\;$ and $\;\gamma_i\ll 1\;$.

The influence of noise, according to Eq. (40), is exponentially small at the 
initial stage. Thus, even a strong noise would not kill the effect, 
although it, of course, would shorten the negative--fluctuation lifetime. 
So, the condition favoring the longer lifetime is weak noise, when 
$\;\gamma_{ij}\ll 1\;$.

After understanding the necessary and favoring conditions for the 
transient effect of a negative--current fluctuation, let us elucidate 
sufficient conditions for the inequality
\begin{equation}
J(t) < 0
\end{equation}
occurring at $\;t\ll 1\;$. As a limiting case we may take $\;t=0\;$ and 
the maximally nonuniform initial density
\begin{equation}
\rho_i(x,0) = f_i(x) = Q_i\delta(x-a_i)
\end{equation}
following from condition (27) under $\;b_i\rightarrow 0\;$. Then from 
expression (20) we readily get \begin{equation}
J(0)=\mu_1Q_1E(a_1,0)+\mu_2Q_2E(a_2,0) .
\end{equation}
This emphasizes once again that for such a nonuniform initial charge 
density the diffusion, relaxation, and noise do not influence much the
value 
of the electric current $\;J(0)\;$. The corresponding electric field, 
defined by Eq. (14), is
\begin{equation}
E(x,0) = 1+4\pi Q_1 [ a_1 -\Theta(a_1 -x)] +4\pi Q_2 [ a_2 -\Theta(a_2 -x) ] ,
\end{equation}
where $\;\Theta(x)\;$ is the unit step function.

Combining Eqs. (41), (43), and (44), we have the inequality
$$ \mu_1Q_1\left\{ 1 +4\pi Q_1\left ( a_1 -\frac{1}{2}\right ) +
4\pi Q_2\left [ a_2 -\Theta(a_2- a_1)\right ]\right\} + $$
\begin{equation}
+ \mu_2Q_2\left\{ 1 +4\pi Q_2\left ( a_2 -\frac{1}{2}\right ) +
4\pi Q_1\left [ a_1 -\Theta(a_1- a_2)\right ]\right\} < 0 
\end{equation}
for the parameters of the system allowing the appearance of a negative 
current.

There can be a number of different cases satisfying inequality (45). To show 
that such situations do really exist, consider a particular example when 
$\;a_1=a_2\equiv a\;$. Then Eq.(41) reduces to
\begin{equation}
(\mu_1Q_1 +\mu_2Q_2)E(a,0) < 0 .
\end{equation}
From equality (44) we get
$$ E(a,0) = 1+4\pi Q \left ( a -\frac{1}{2}\right ) \qquad 
(Q \equiv Q_1 +Q_2 ) . $$
Recall that $\;\mu_1\;$ and $\;Q_1\;$ are positive, while $\;\mu_2\;$ and 
$\;Q_2\;$ are negative; so that $\;\mu_iQ_i > 0\;$. Thence, inequality 
(46) can be hold only if
\begin{equation}
E(a,0) < 0 .
\end{equation}
As follows from solution (28), the quantity $\;\mu_iE\;$ plays the role of 
the effective velocity of motion for the corresponding charge packet. In the 
case of inequality (47), we have $\;\mu_1E < 0\;$ and $\;\mu_2E > 0\;$. This 
means that the positive carriers effectively move against the applied voltage; 
and the negative carriers, along the latter; that is, they move oppositely to 
what one would expect. Hence, the negative electric current is related to the 
anomalous drift of charge carriers.

Substituting into Eq. (47) the electric field, we find 
\begin{equation}
4\pi Q \left ( \frac{1}{2} - a\right ) > 1 .
\end{equation}
Depending on whether $\;Q\;$ is positive or negative, inequality (48) yields
$$ a < \frac{1}{2} -\frac{1}{4\pi Q} \qquad ( Q > 0) , $$
\begin{equation}
a > \frac{1}{2} +\frac{1}{4\pi|Q|} \qquad (Q < 0) .
\end{equation}
Taking also into account that $\;0<a<1\;$, we obtain from inequalities (49) 
the condition 
\begin{equation}
|Q|> \frac{1}{2\pi} .
\end{equation}
Equations (49) and (50) are sufficient conditions for the appearance of a 
negative electric current at the initial stage of the process. Similarly,
it is easy to show that the currents (17) and (18) can also become
negative in the transient regime.

To study in more detail the behaviour of the electric current as a 
function of time, we have solved Eqs.(9) and (10) numerically. In 
accordance with the above analysis, the case favoring the 
negative--current fluctuation is accepted, when $\;\gamma_{ij}\ll 1\;$. 
The initial conditions are given by the Gaussian form (27). The voltage 
integral (12) plays the role of the boundary condition for the electric 
field. For the charge densities one may take the Neumann or Dirichlet 
boundary conditions [1]. We have tried both and found that the general 
picture does not change much, with the only difference that the 
calculational procedure is less stable for the Dirichlet conditions. To 
achieve the best stability, we opted for the Neumann boundary conditions.

For the characteristic parameters we accept the values typical of
semiconductors [1,2], such as $\;Si\;$. Then the diffusion coefficients
are $\;D_1\sim 10\; cm^2/s,\; D_2\sim 30\; cm^2/s$. The mobility of
positive carriers $\mu_1\sim 500\; cm^2/Vs$ for the average concentration
$\;10^{13}-10^{14}cm^{-3}\;$ and $\;\mu_1\sim 200\; cm^2/Vs\;$ for the
concentration $\;10^{17}-10^{18}cm^{-3}\;$. The mobility of electrons
$\mu_2\sim 1500\;$ for the average concentration $\;10^{13}-10^{14}\;$ and  
$\;\mu_2\sim 400\;$ for the concentration $\;10^{16}-10^{18}\;$. The
recombination time $\;\tau_1\sim\tau_2\sim 10^{-12}-10^{-10}s\;$, hence
the relaxation width $\;\gamma_1\sim\gamma_2\sim 10^{10}-10^{12}s^{-1}\;$. We
consider a plane device of the size $\;A\sim 1\;cm^2,\; L\sim0.1-1\;cm\;$,
with the applied voltage $\;V_0\sim 10^3-10^5\; V\;$. For the calibration
parameters in Eq.(8) we get $\mu\sim 10^3cm^2/Vs,\;V_0\sim 10^3-10^5V,\;
D_0\sim 10^6-10^8cm^2/s,\; E_0\sim 10^3-10^5V/cm,\; Q_0\sim 10^3-10^6V\;cm,
\; \rho_0\sim 10^3-10^7V/cm^2,\; J_0\sim 10^8-10^{15}V\; cm^2/s$. The
transit time (6) is $\tau_0\sim 10^{-6}-10^{-9}s$.

The results of our numerical calculations are presented in the Figs.1 to
6, where we show the time dependence of the electric current across the
boundaries as well as the behaviour of the total current through the
device. These figures demonstrate that the appearance of negative electric
current is really a transient effect occurring at dimensionless times
$t\ll 1$, which in dimensional units means that $t\ll\tau_0$. All values
in the figure captions are given in dimensionless units employing the
calibration parameters from Eq. (8). Also, for shortness, we write
$a_1\equiv a$ and $b_1=b$.

Fig.1 shows that the electric current through the left boundary of the 
semiconductor sample, through its right boundary, and the total electric 
current are different. This difference is not merely quantative but can 
be qualitative, so that the negative current may happen at the right 
surface and on average through the sample, but may be absent on the left 
surface. Such a difference depends on semiconductor characteristics as
mobilities and relaxation widths. This suggests the possibility of 
employing the principal difference in the behaviour of the currents for 
extracting information on the semiconductor characteristics. For example, 
in Fig. 2 it is seen that changing the electron mobility mainly 
influences the current across the left surface. The transient negative 
current becomes more pronounced when increasing the absolute value of the 
electron mobility, as is seen in Fig. 3. The lifetime of the 
negative--current fluctuation strongly depends on the relaxation width, 
which is illustrated in Fig. 4. Increasing the relaxation width shortens 
the fluctuation lifetime. Figure 5 demonstrates the role of the total 
initial charge on the occurrence of the negative current, and Fig. 6 shows 
the role of the initial distribution of charge carriers. The importance 
of special conditions for the initial charge and its location has been
discussed in detail above. The amplitude of the transient negative--current 
fluctuation becomes smaller when the charge layer at initial time is 
shifted farther from the left surface of the semiconductor sample.

\section{Discussion}

We have considered electric processes in nonequilibrium nonuniform 
semiconductors. The transport equations are taken  in the standard 
drift--diffusion approximation that is widely used for describing realistic 
semiconductor devices. Both analytical and numerical solutions of these 
equations are accomplished.

It is shown that under special circumstances an unusual transient 
phenomenon appears displaying negative electric current. The 
necessary condition for such an anomalous current is nonuniformity of 
carrier densities at the initial stage. A general sufficient condition 
(45) is derived and its particular forms (49) and (50) are analysed in 
detail. We studied the influence of diffusion, relaxation, and of 
generation--recombination noise and showed that these processes, even 
being strong, do not destroy the effect although may shorten the lifetime 
of a negative--current fluctuation. Therefore such an anomalous electric 
current can really be observed in semiconductor devices.

An important physical question is how one could use the considered effect 
for practical applications. Several possibilities of using this effect 
can be suggested:

\vspace{2mm}

(i) The appearance of the transient negative--current fluctuation is 
rather sensitive to the characteristic parameters of semiconductor, such 
as the carrier mobilities and relaxation widths. Therefore one could use 
the observation of the current fluctuation for defining these parameters. 
This could be done in the following way. Assume that we know all 
parameters except one, say a mobility or a relaxation coefficient. 
Comparing the time--dependence of the measured current with that of the 
calculated one, we may try to find such a value of the sought parameter 
that the measured and the calculated behaviour of the electric currents 
be as close as possible to each other, in the optimal case, be almost 
coinciding. Then the so fitted quantity would give the value of the 
sought parameter.

\vspace{2mm}

(ii) When the layer of charge carriers is formed by irradiating 
semiconductor with an ion beam, the stuck ions are distributed 
approximately in the Gaussian law centered at the mean free path of the 
ions. A necessary condition for the occurrence of the negative current is 
that the charge layer is located at a particular distance from the 
semiconductor surface, as e.g. in Eq. (49). Thence, this effect is very 
sensitive to the initial location of charge carriers and, thus, could be 
used for measuring the mean free path of ions in specific semiconductor 
materials.

\vspace{2mm}

(iii) The value of the total charge in the initial nonuniform layer is 
also crucially important for the occurrence of the negative current, as 
is seen from Eqs. (48)--(50). Hence, studying this current, we could 
measure the initial charge. The latter may be unknown when the initial 
distribution of charge carriers is formed by irradiating semiconductor 
with narrow laser beams whose influence on the generation of carriers in 
not precisely known.

\vspace{2mm}

(iv) Semiconductor devices often work in the close vicinity of radiation 
sources, such as atomic reactors, or under the influence of other strong 
radiation, as cosmic rays. In the presence of radiation, the functioning 
of semiconductor devices can be drastically disturbed because of the 
arising carrier nonuniformities. This can lead not only to the 
malfunctioning of semiconductor devices but even to dramatic accidents. 
In order to prevent from these, one could employ controlling schemes 
reacting to the appearance of the negative current, signaling by this 
that the level of the carrier nonuniformity induced by irradiation has 
become dangerous.

\vspace{2mm}

(v) As we have shown, the generation--recombination noise does not 
destroy the effect of the negative--current fluctuation. However, there
exist other types of noise [7--9] whose influence on the supression of 
this effect can be different. Therefore, analysing the peculiarity
of the electric negative--current fluctuation, one could judge what kind 
of noise dominates the process in the studied semiconductor.

It is certainly not possible to enumerate all feasible applications of 
the considered effect. But we hope that the examples listed above do 
demonstrate that the specific unusual features of the negative--current 
transient effect could provide us several interesting physical 
applications. Three types of such applications are, generally, 
admissible. One type is for investigating the characteristics of 
semiconductor materials. Another type can be used for studying the 
properties of irradiating beams. And, finally, this effect can be 
employed for the practical purpose of creating special controlling 
instruments.

\vspace{5mm}

{\bf Acknowledgement}

\vspace{3mm}

We appreciate a grant from the University of Western Ontario, London,
Canada. One of the authors (M.S.) is grateful to NSERC of Canada for
financial support in the form of a research grant.

\newpage

{\Large{\bf Appendix. Stationary Solution}}

\vspace{5mm}

Here we prove that Eqs.(33) and (34) are the stable stationary solutions 
of the exact equations (9) and (10) under the voltage condition (12). To 
this end, because of equality (32), it is sufficient to prove that
$$ \lim_{t\rightarrow\infty}\rho_i^{reg}(x,t)= 0 , \qquad 
\lim_{t\rightarrow\infty}E^{reg}(x,t) = 1 , $$
where $\;E^{reg}\;$ implies the functional (14) including the dependence 
on only $\;\rho_i^{reg}\;$. In what follows we shall write for brevity 
$\;\rho_i\;$ and $\;E\;$ keeping in mind $\;\rho_i^{reg}\;$ and 
$\;E^{reg}\;$. The proof will be based on the method of multipliers [15].

Write Eq.(9) in the form
$$ \frac{\partial}{\partial t}\rho_i(x,t) =v_i(x,\rho,t) , $$
with the velocity field
$$ v_i(x,\rho,t) = D_i\frac{\partial^2\rho_i}{\partial x^2} 
-\mu_i\frac{\partial}{\partial x}(\rho_iE) -\gamma_i\rho_i . $$
Define the multiplier matrix
$$ M_{ij}(x,x',t) =\frac{\delta\rho_i(x,t)}{\delta\rho_j(x',0)} $$
and the Jacobian matrix
$$ L_{ij}(x,x',\rho,t) =\frac{\delta v_i(x,\rho,t)}{\delta\rho_j(x',t)} . $$
The latter, with the given velocity field, consists of the elements
$$ L_{ii}(x,x',\rho,t) =\left ( D_i\frac{\partial^2}{\partial x^2} 
-\mu_iE\frac{\partial}{\partial x} -\gamma_i\right )\delta(x-x') - $$
$$ - 4\pi\mu_i(2\rho_i+\rho_j)\delta(x-x') -
\mu_i\frac{\partial\rho_i}{\partial x}\frac{\delta E(x,t)}{\delta\rho_i(x',t)} , $$
$$ L_{ij}(x,x',\rho,t) = -4\pi\mu_i\rho_i\delta(x-x') -
\mu_i\frac{\partial\rho_i}{\partial x}\frac{\delta E(x,t)}{\delta\rho_j(x',t)} , $$
where $\;i\neq j\;$. Here, the variational derivative of the electric 
field can be found from expression (14) giving
$$ \frac{\delta E(x,t)}{\delta\rho_i(x',t)} = 4\pi\left [
\frac{\delta Q(x,t)}{\delta\rho_i(x',t)} + x' - 1\right ] , \qquad
\frac{\delta Q(x,t)}{\delta\rho_i(x',t)} =\Theta(x-x') . $$
Varying the evolution equation for $\;\rho_i\;$, we obtain the equation
$$ \frac{\partial}{\partial t}M_{ij}(x,x',t) =\sum_k\int_0^1
L_{ik}(x,x'',\rho,t)M_{kj}(x'',x',t)dx'' $$
for the multiplier matrix, with the initial condition
$$ M_{ij}(x,x',0) =\delta_{ij}\delta(x-x') , $$
following from the variation of condition (13).

For the case $\;\rho_i=0\;$ and $\;E=1\;$, the Jacobian matrix is
$$ L_{ij}(x,x',0,t) =\delta_{ij}\left (
D_i\frac{\partial^2}{\partial x^2} -\mu_i\frac{\partial}{\partial x} -
\gamma_i\right ) \delta(x-x') . $$
This leads to the equation
$$ \frac{\partial M_{ij}}{\partial t} =
\left ( D_i\frac{\partial^2}{\partial x^2} 
-\mu_i\frac{\partial}{\partial x} -\gamma_i\right ) M_{ij}  $$
for the multiplier matrix. From the latter equation, invoking the Fourier 
transform
$$ M_{ij}(x,x',t) 
=\frac{1}{2\pi}\int_{-\infty}^{+\infty}M_{ij}(k,t)e^{ik(x-x')}dk , $$
we find
$$ M_{ij}(k,t) =\delta_{ij}\exp\left\{ -(i\mu_ik + D_ik^2 +\gamma_i)t
\right\} $$

As is seen,
$$ | M_{ij}(k,t)| < 1 $$
for all $\;k\in(-\infty,+\infty)\;$ and $\;t>0\;$. Hence the motion in 
the vicinity of the stationary solutions $\;\rho_i=0\;$ and $\;E=1\;$ is 
stable. It is also asymptotically stable since
$$ \lim_{t\rightarrow\infty} M_{ij}(k,t) = 0 $$
for all $\;k\in(-\infty,+\infty)\;$. This completes the proof.

\newpage

\begin{center}
{\bf Figure Captions}
\end{center}

{\bf Fig.1} The time behaviour of the electric current (17) across the
left boundary (solid line), of the current (18) across the right boundary
(long--dashed line), and of the total current (19) through the
semiconductor device (short--dashed line). The used parameters, in
dimensionless units, are $Q_1=1,\; Q_2=-\frac{1}{2},\;\mu_1=1,\;\mu_2=-10,
\;\gamma_1=\gamma_2=10,\; a=0.25,\; b=0.1,\; D_1 = 10^{-3},\; D_2=3D_1$.

\vspace{5mm}

{\bf Fig.2} The same as in Fig.1, but for $\mu_2=-3$ and $a=0.05$. The
smaller absolute value of the electron mobility influences mainly the
current across the left boundary.

\vspace{5mm}

{\bf Fig.3} Electric current (19) through the semiconductor device as a
function of time for different mobilities. The fixed parameters are
$Q_1=1,\; Q_2=-\frac{1}{2},\;\gamma_1=\gamma_2=1,\; a=0.25,\; b=0.1,\;
D_1=10^{-3},\; D_2=3D_a,\;\mu_1=1$. The curves correspond to the varying
mobility: $\mu_2=-10$ (solid line), $\mu_2=-5$ (long--dashed line), and
$\mu_2=-3$ (short--dashed line). Increasing the absolute value of the
electron mobility makes the negative current more pronounced.

\vspace{5mm}

{\bf Fig.4} The same as in Fig.3, except for $\mu_2=-3$ and for varying
relaxation widths: $\gamma_1=\gamma_2=25$ (solid line), $\gamma_1=\gamma_2=10$
(long--dashed line), and $\gamma_1=\gamma_2=1$ (short--dashed line). Increasing
relaxation width suppresses the negative current.

\vspace{5mm}

{\bf Fig.5} The time dependance of the electric current (19) for the
following parameters: $\mu_1=1,\;\mu_2=-3,\;\gamma_1=\gamma_2=1,\; a=0.25,\;
b=0.1,\; D_1=10^{-3},\; D_2=3D_1,\; Q_1=1$, and for varying $Q_2=0$
(solid line), $Q_2=-\frac{1}{4}$ (long--dashed line), $Q_2=-\frac{1}{2}$
(short--dashed line), $Q_2=-\frac{3}{4}$ (dotted line), and $Q_2=-1$
(dash--dotted line).

\vspace{5mm}

{\bf Fig.6} The same as in Fig. 5, but for the fixed $Q_2=-\frac{1}{4}$
and for a varying location of the initial distribution of carriers:
$a=0.1$ (solid line), $a=0.2$ (long--dashed line), $a=0.3$ (short--dashed
line), $a=0.4$ (dotted line), and $a=0.5$ (dash--dotted line).

\end{document}